\documentclass[global,twocolumn]{svjour}
\usepackage{graphicx}
\usepackage{color}
\usepackage{times}
\usepackage{amsmath}
\usepackage{amssymb}

\newcommand{\be}{\begin{equation}}
\newcommand{\ee}{\end{equation}}

\newcommand\rpict[1]{\ref{fig:#1}}

\newcommand{\eq}[1]{Eq.~(\ref{eq:#1})}

\newcounter{Fig}

\newcommand{\f}{\text{f}}
\newcommand{\sinc}{\text{sinc}}
\newcommand{\LAM}{\Lambda}
\newcommand{\lAver}{l_m}
\newcommand{\aver}[1]{\left<#1\right>}

\newcommand{\indSF}{\text{\tiny SF}}
\newcommand{\indFH}{1}
\newcommand{\indSH}{2}
\newcommand{\indTH}{3}

\newcommand{\lamFH}{\lambda}

\newcommand{\Electricfield}[1]{E_{#1}}
\newcommand{\ESF}{\Electricfield{\indSF}}
\newcommand{\EFH}{\Electricfield{\indFH}}
\newcommand{\ESH}{\Electricfield{\indSH}}
\newcommand{\ETH}{\Electricfield{\indTH}}

\newcommand{\Intensity}[1]{I_{#1}}
\newcommand{\ISF}{\Intensity{\indSF}}
\newcommand{\IZero}{\Intensity{0}}

\newcommand{\ISH}{\Intensity{\indSH}}
\newcommand{\ITH}{\Intensity{\indTH}}

\newcommand{\wavevector}[1]{k_{#1}}
\newcommand{\kFH}{\wavevector{\indFH}}
\newcommand{\kSH}{\wavevector{\indSH}}
\newcommand{\kTH}{\wavevector{\indTH}}

\newcommand{\wavevectorVec}[1]{\vec{k}_{#1}}
\newcommand{\kFHvec}{\wavevectorVec{\indFH}}
\newcommand{\kSHvec}{\wavevectorVec{\indSH}}
\newcommand{\kTHvec}{\wavevectorVec{\indTH}}

\newcommand{\Dk}[1]{{q_{#1}}}

\newcommand{\DkSH}{\Dk{\indSH}}
\newcommand{\DkTH}{\Dk{\indTH}}

\newcommand{\Dkvec}[1]{\vec{q}_{#1}}

\newcommand{\DkSHvec}{\Dkvec{\indSH}}
\newcommand{\DkTHvec}{\Dkvec{\indTH}}

\newcommand{\M}[1]{M(#1)}
\newcommand{\MSq}[1]{|\M{#1}|^2}
\newcommand{\MSqAver}[1]{\left<\MSq{#1}\right>}

\newcommand{\CharacteristiFunction}[1]{\varphi(#1)}
\newcommand{\ChDk}[1]{\CharacteristiFunction{\Dk{#1}}}

\newcommand{\emissionAngle}[1]{\alpha_{#1}}

\newcommand{\eaSH}{\emissionAngle{\indSH}}
\newcommand{\eaTH}{\emissionAngle{\indTH}}

\journalname{Applied Physics B: Lasers and Optics}
\begin{document}

\title{Parametric wave interaction in  one-dimensional nonlinear photonic crystal with randomized distribution of second-order nonlinearity}

\author{K. Kalinowski\inst{1,2}\and V. Roppo\inst{1,3}\and T. {\L}ukasiewicz\inst{4}\and M. \'{S}wirkowicz\inst{4}\and
Y. Sheng\inst{1}\and and W. Krolikowski\inst{1}}

\institute{
Laser Physics Center, Research School of Physics and Engineering, Australian National University, Canberra, ACT 0200, Australia, \email{xkk124@physics.anu.edu.au}\and
Nonlinear  Physics Center, Research School of Physics and Engineering, Australian National University, Canberra, ACT 0200, Australia\and
Departament de Fisica i Enginyeria Nuclear, Escola Tecnica Superior d'Enginyeries Industrial y Aeronautica de Terrassa, Universitat Politecnica de Catalunya, Rambla Sant Nebridi, 08222 Terrassa, Barcelona, Spain\and
Institute of Electronic Materials Technology, W\'{o}lczy\'{n}ska 133, 01-919 Warsaw, Poland
}
\maketitle
\begin{abstract}
We theoretically study the parametric wave interaction in nonlinear optical media with randomized distribution of the quadratic nonlinearity $\chi^{(2)}$. In particular, we discuss properties of second and cascaded third harmonic generation. We derive analytical formulas describing emission properties of such harmonics in the presence of $\chi^{(2)}$ disorder and show that the latter process is governed by the characteristics of the constituent processes, i.e. second harmonic generation and sum frequency mixing. We demonstrate the role of randomness on various second and third harmonic generation regimes such as Raman-Nath and \v{C}erenkov nonlinear diffraction. We show that  the randomness-induced incoherence in the wave interaction leads to deterioration of conversion efficiency  and angular spreading of harmonic generated in the processes relying on  transverse phase matching such as Raman-Nath interaction. On the other hand, the \v{C}erenkov  frequency generation is basically insensitive to the domain randomness.
\end{abstract}




\section{Introduction}
Nonlinear photonic crystals (NLPC), i.e.  periodically poled ferroelectric crystals or orientation-patterned semiconductors with quadratic nonlinearity, have been commonly used to realize efficient frequency conversion. The spatially periodic modulation of the sign of $\chi^{(2)}$ nonlinearity ensures the synchronization of the phases of interacting waves via the so called quasi phase matching (QPM)~\cite{Armstrong:PhysRev:1962,Franken:rmp:1963,Fejer:jqe:1992}. In the simplest case of one-dimensional (1D) periodic NLPC [Fig.1(a-b)] and collinear second harmonic generation (SHG) high conversion efficiency is assured by choosing the period of nonlinearity modulation $\Lambda$ such that the phase mismatch $\Dk{}$ satisfies the following relation $\Dk{}= |k_2-2k_1| = 2\pi/\Lambda$, where $k_1$, $k_2$ are wave vectors of the fundamental and second harmonic, respectively.

The QPM technique is so versatile that by modulating spatially the quadratic nonlinearity and utilizing the so-called non-collinear (or transverse) type of wave interaction one can also spatially shape the wavefront of generated waves as has been demonstrated in case of Bessel and Airy beams SHG~\cite{Saltiel:oe:2007,Ellenbogen:NatPhot:2009}. However, because of resonant character of the QPM, perfectly periodic structure is efficient only for a particular choice of the wavelengths of interacting waves.

Few methods have been proposed in order to extend the operational bandwidth of the QPM technique. They all rely on engineered nonlinear photonic structures and involve, for instance, structures consisting of periodically poled sections with different periods or multi-period and quasi-periodic structures, i.e. nonlinear superlattices~\cite{Zhu:Science:1997,Lu:jnopm:2007,Sapaev:oe:2008,Fradkin-Kashi:prl:2001,Arie:LPOR:2010}.
\begin{figure}[ht]
\centerline{\includegraphics[width=8.5cm]{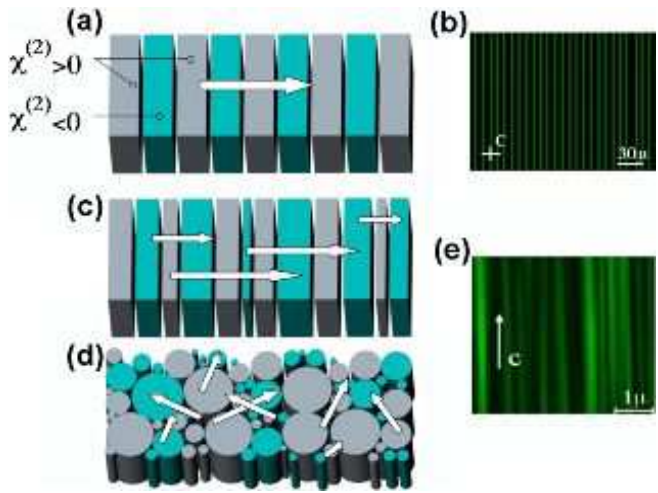}}
\caption{(color online) (a) Illustrating the concept of periodically poled NLPC. Arrow indicates the reciprocal lattice vector (RLV). (b) Domain structure in commercial sample of periodically poled lithium niobate crystal with nominal poling period of 14 $\mu$m (visualized  via  the second harmonic nonlinear microscopy~\cite{Sheng:oe:2010}). (c-d) Schematic representation of the  NLPC with (c) one- and (d) two-dimensional  random spatial domain distribution. Arrows represent various RLV. (e) Random ferroelectric domain pattern in  SBN crystal (visualized  via  SH nonlinear microscopy~\cite{Sheng:oe:2010}). ``C'' in (b) and (e) indicates the direction of the optical axis.}
\end{figure}
\begin{figure}[ht]
\centerline{\includegraphics[height=4cm]{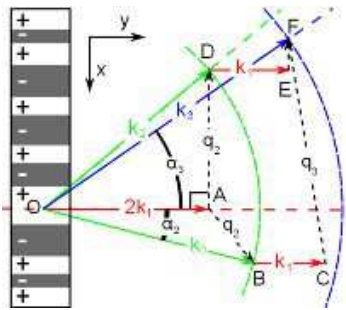}}
\caption{(color online) Illustrating the phase matching conditions for the second and third harmonic generation via noncollinear interaction in 1D nonlinear photonic crystal with randomized distribution of second order nonlinearity. $k_2$ and $k_3$ represent wave vectors of the second and third harmonics, respectively. Green and blue rings (with corresponding radii $k_2$ and $k_3$) define all possible emission directions of the second and third harmonics.}
\label{fig:Diagram}
\end{figure}
Another possibility offers media with disordered or randomized distribution of the sign of nonlinearity [see Fig.1(c)]. It is known that some as-grown ferroelectric crystals, such as strontium barium niobate (SBN) [Fig.1(d-e)]  or calcium barium niobate, consist of elongated ferroelectric domains oriented along its polar axis having random distribution of their size and orientation~\cite{Sheng:oe:2010,PhysRevB.77.054105,molina:071111}. As such they represent disordered nonlinear photonic crystal, which can be considered as being composed of an infinite number of periodic structure, corresponding to a pool of Reciprocal Lattice Vectors (RLV) which ensure the phase matching for the harmonic emissions at any incident wavelength in its transmission band.

The extensive studies of the second-order nonlinear processes in disordered NLPC resulted in demonstration of a number of effects including e.g.,  angular broadening of the emitted harmonics~\cite{Roppo:10,Ayoub:11}, multi-frequency conversion~\cite{Romero:apl:2002,Romero:jap:2003,Ramirez:apb:2005}, conical and planar (non-collinear) frequency emission~\cite{Tunyagi:prl:2003,Trull:oe:07,Roppo:oe:2008} and its application in realization of optical autocorrelator for short pulse diagnostics and monitoring~\cite{Fischer:apl:2006,Fischer:apl:2007}. Few recent works have demonstrated beneficial effect of controlled randomization of originally periodic pattern on second harmonic generation~\cite{Sheng:apl:11:broadband,Varon:ol:early:11}. It is worth to add that the efficiency of the conversion process in those randomized media critically depends on degree of the nonlinearity  disorder. It is as small as a 0.001\% in completely  random crystals~\cite{molina:071111}  but it can be as high as tens of a percent for weakly randomized media~\cite{Sheng:apl:11:broadband}.

It has been shown recently in experiments with as-grown SBN crystal that under certain conditions the ferroelectric crystal with random domain distribution can be used for broadband cascaded third harmonic generation~\cite{Molina:afm:2008,Wang:jpb:2010,Wang:oe:09,Sheng:ol:09,Sheng:apb:11}. In such case the randomness of the nonlinearity contributes to both constituent processes: second harmonic and sum frequency generation. Subsequently, as the experiments showed, the  spatial light intensity distribution of the emitted third and second harmonics are significantly different.

The theoretical studies of the wave interaction in randomized media deal with either transverse or collinear type of SHG depending on whether the generated harmonics are  emitted collinearly or transversely with respect to the propagation direction of the fundamental beam. The transversely phase matched SHG has been considered  already  in 1972 by Dolino {\em et al.} who used Green function approach to investigate  the role of randomness in domain shape on second harmonic generation~\cite{Dolino:prb:1970}.
In their  2001 work Le Grand {\em et al.}   analyzed  the role  of random one-dimensional domain distribution on the angular distribution  of the second harmonics. These authors also considered the effect  of various statistical models of randomness on harmonic emission~\cite{LeGrand:oe:2001}.

Spatial properties of the second harmonic in random medium and the effect of group velocity dispersion have been investigated  recently by Shutov {\em et al.}~\cite{Shutov:os:08}.
In their  work Tunyagi {\em et al.}~\cite{Tunyagi:prl:2003} used numerical simulations in calculating second harmonic field emitted transversely from 1-D random crystal. Numerical simulations were also used by Vidal and  Martorell~\cite{Vidal:prl:2006} to study the diffusive character of the second harmonic generated in random 1D crystal. These simulations also confirmed the well known analytical results  that the disorder-induced loss  of phase correlation leads to the emitted signal growing  linearly with propagation distance~\cite{Morozov:jetpl:2001,Morozov:qe:2004}. In their recent work Bravo-Abad {\em et al.} suggested that light scattering by randomly distributed ferroelectric  domains might contribute towards appearance of forward emission peak of the second harmonic signal~\cite{Bravo-Abad:oe:2010}. The  spatial intensity distribution of the generated harmonics in random quadratic media and its application for the diagnostics of disorder in nonlinear structures have been discussed in papers  by Roppo {\em et al.}~\cite{Roppo:10} and Pasquazi {\em et al.}~\cite{Pasquazi:pj:2010}.

As far as the collinear interaction is concerned S. Helmfrid and G. Arvidsson~\cite{Helmfrid:91}  in their 1991 paper analyzed the influence of randomly varying domain length  on the efficiency of the  second harmonic generation. They demonstrated that randomness, while degrading  the conversion efficiency, increases the operational bandwidth. Recently Pelc {\em et al.}  considered the same interaction geometry to demonstrate efficiency enhancement away from the QPM peak~\cite{Pelc:11:0l}. Extensive theoretical investigation of the random perturbation of the ideal periodically poled structure in ferroelectrics on collinear frequency conversion has been  conducted  by Fejer {\em et al.}~\cite{Fejer:jqe:1992}. It should be also mentioned that the randomness can be introduced inadvertently in the  fabrication of QPM structures via periodic poling [see Fig.1(b)]. This typically has an adverse consequences on the intended performance of the QPM device leading, e.g. to decreased frequency conversion efficiency~\cite{Fejer:jqe:1992,Helmfrid:91,Pelc:11:0l,Pelc:10:ol,Stivala:ol:10}.


In this work we present extensive theoretical studies of the non-collinear parametric wave interaction in crystal with randomized one-dimensional distribution of quadratic nonlinearity. In particular, we investigate for the first time the spatial properties of light emissions in the cascaded process of third harmonic generation in the randomized nonlinear optical structures. Using the exact analytical formulas for the intensity of generated waves obtained in the regime of strong fundamental beam and plane wave approximation we compare the emission characteristics of second and third harmonics as a function of the degree of disorder and the wavelength.

Our theoretical analysis uses the plane wave assumption for interacting fields. While being obviously restrictive  the fully analytical character of this approach allows us to obtain simple physical picture of all relevant mechanisms involved in the frequency conversion in random nonlinear photonic crystals.
\section{Sum Frequency Mixing}
Let us consider interaction of two optical plane waves with amplitudes $E_a,E_b$ with the frequencies $\omega_a$ and $\omega_b$ in nonlinear photonic crystal with the spatially modulated sign of nonlinear coefficient. The nonlinearity will lead, among others, to the generation of the third wave $\ESF$ at the frequency $\omega_{\indSF}=\omega_a+\omega_b$ whose amplitude can be expressed as~\cite{Russel:jqe:2001}:
\begin{equation}\label{eq:ESF_1}
  \ESF(\Dkvec{}) \propto i\frac{d\,E_a E_b}{L_x} \int\int_A \M{\vec{r}} e^{i\Dkvec{}\vec{r}}d\vec{r},
\end{equation}
where $d$ is the nonlinearity coefficient ($d=\chi^{(2)}/2$) and integration is performed over the whole area $A=L_xL_y$ of the nonlinear crystal, $L_x$, $L_y$ are the lengths of the crystal in $x$, $y$ directions and $\M{\vec{r}}=\pm1$ represents the modulation of the sign of nonlinear coefficient. The phase mismatch vector is $\Dkvec{}=\vec{k}_{\indSF} - \vec{k}_a-\vec{k}_b$ with $k_i=n_i2\pi/\lambda_i $ representing wave vector and $n_i$ is refractive index of the corresponding wave. In this paper we consider one-dimensional structure with nonlinearity modulation only in x direction (Fig.~\ref{fig:Diagram}) such that:
\begin{equation}\label{eq:SpatialModulation}
  \M{\vec{r}} = \sum_{j=0}^{N-1} (-1)^j \Pi_{x_j,x_{j+1}}(x),
\end{equation}
where $x_j$ is a position of $j$-th domain wall, $N$ is number of domains and $\Pi_{x_j,x_{j+1}}(x)=1$ for $x_{j}\le x\le x_{j+1}$ and $0$ otherwise. In this case integral (\ref{eq:ESF_1}) can be separated into $x$ and $y$ directions. Additionally, because M(x)=0 outside crystal ($x<x_0$ or $x>x_N$) integration over x can be extended into infinity and one gets:
\begin{equation}\label{eq:ESF_2}
  \begin{split}
    \ESF(\Dkvec{}) &\propto i \frac{d E_a E_b}{L_x} \int_{0}^{L_y} e^{i\Dk{}_y y}dy\int_{-\infty}^{\infty} \M{x} e^{ i\Dk{}_x x}dx \\
                   &\propto i \frac{d E_a E_b L_y}{L_x} \,\sinc\left(\frac{\Dk{}_y L_y}{2}\right)\,\M{\Dk{}_x},
  \end{split}
\end{equation}
where $M(\Dk{}_x)$ denotes a Fourier transform of spatially modulated nonlinear coefficient:
\begin{equation}\label{eq:MDk}
    \M{\Dk{}_x} = \frac{i}{\Dk{}_x}\left(2\sum_{j=1}^{N}(-1)^je^{i\Dk{}_x x_j}-(-1)^Ne^{i\Dk{}_x x_N}+1\right).
\end{equation}
In general the last two terms in the bracket can be neglected~\cite{Fejer:jqe:1992} what results in:
\begin{equation}\label{eq:MDkSimplified}
    \M{\Dk{}_x} = \frac{2i}{\Dk{}_x}\sum_{j=1}^{N}(-1)^j\prod_{k=1}^j e^{i\Dk{}_x l_k},
\end{equation}
where $l_k=x_k-x_{k-1}$ is the width of $k$-th domain.

In case of perfectly periodic structure where all the nonlinear domains are equal $l_k=\LAM/2$ ($\LAM$ is a nonlinearity modulation or poling period) the strongest Fourier coefficient $M(\Dk{}_x)$ corresponds to phase mismatch vector equal to the first RLV in which case $\M{\Dk{}_x=2\pi/\LAM} = i\LAM N/\pi=2iL_x/\pi$. If the domains' widths $l_k$ are not equal but are subject to randomness, then one has to take into account their statistical properties. This means that the ensemble averaging should be used to describe  the emitted light intensity.  Additionally, in real settings when domains are actually redistributed in all three dimensions the emitted harmonic contains  contributions originating in different locations and hence the measured intensity is, to certain extent, represented by an average over different realization $\aver{\ISF} \propto \aver{|\ESF|^2}$:
\begin{equation}\label{eq:ISF}
  \aver{\ISF(\Dkvec{})} \propto \frac{1}{L_x^2}\sinc^2\left(\frac{\Dk{}_y L_y}{2}\right)\,\MSqAver{\Dk{}_x},
\end{equation}
where $\aver{\,\cdot\,}$ denotes averaging over domains width $l$. The expression $\MSqAver{\Dk{x}}$ has been discussed before~\cite{Fejer:jqe:1992,LeGrand:oe:2001}. We adopted here the approach as described by Le Grand {\em et al.} with some modifications that allow us to extend the  results of SHG to SFM and subsequently to the cascaded THG.

If domains width varies randomly its statistics can be described by a probability density function $w(l)$ with characteristic function $\CharacteristiFunction{\Dk{}}=\int w(l)\exp(i\Dk{} l)dl$. Then for $|\ChDk{}|<1$ and large number of domains ($N\gg1$) one obtains:
\begin{equation}\label{eq:MDkAvG}
    \MSqAver{\Dk{}_x} = \frac{4N}{\Dk{x}^2}\left( 1 - 2 Re\left(\frac{\ChDk{x}}{\ChDk{x}+1}\right) \right) = \frac{4N}{\Dk{x}^2}\f(\Dk{x})
\end{equation}
what is exactly the formula derived by LeGrand {\em et al.}~\cite{LeGrand:oe:2001}.

In order to calculate the averaged emission intensity $\aver{\ISF}$ one has to choose a function that describes statistical  distribution of the domain widths. A natural choice would be Gaussian distribution.  However, the Gaussian model  is only applicable to rather small values of dispersions $\sigma$  (in comparison to the average domain size $\lAver$) because for dispersions comparable or bigger than $\lAver$ it results in negative values of domain width. The more appropriate choice is therefore the  Gamma distribution which is defined for $l\in(0,\infty)$ as
\begin{equation}
w(l)=\frac{l^{k-1} \theta ^{-k} e^{-\frac{l}{\theta }}}{\Gamma (k)},
\end{equation}
where $k$ and $\theta$ are the so called  shape and scale  parameters, respectively and $\Gamma$ is gamma function.
In this case the characteristic function is $\varphi(q)= (1-i q \theta )^{-k} $. Maximum of the distribution, i.e. the most likely value (or mode) is $\lAver=(k-1)\theta$ while standard deviation is $\sigma=\sqrt{k \theta^2}$. Gamma distribution, unlike the Gaussian model, enables one to account for all possible dispersions regardless of the mode value.
In this paper, for the sake of clarity of the presented results, in all calculations we use Gamma distribution with fixed mode value $\lAver$=1 $\mu$m (what corresponds to poling period $\LAM$=2 $\mu$m) and we will characterize the randomness of the structure by its standard deviation $\sigma$. Also in all calculations we use sample length in $y$ direction $L_y=$500 $\mu$m and a number of domains $N$=500.

\begin{figure}[th]
  \includegraphics[width=8.5cm]{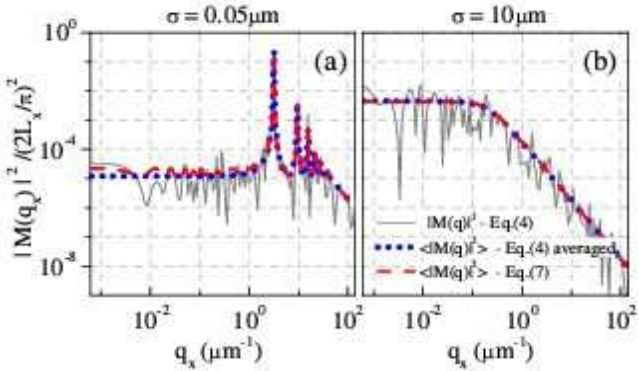}
\caption{(color online) Modulus square of the Fourier coefficient (normalized to $(2L_x/\pi)^2$) of the sum frequency mixing process as a function of the phase mismatch parameter $\Dk{x}$ for an almost  periodic $\sigma$=0.05 $\mu$m (a) and randomized $\sigma$=10 $\mu$m (b) nonlinear photonic structure. Solid (grey) and  dashed (red) lines represent formula \eq{MDk} calculated for particular realization of the nonlinearity (or domain)  distribution  and  averaged over 300 numerically generated random structures, respectively. Dotted (blue) line, which almost exactly overlaps with the dashed red line, represents the simplified formula (\ref{eq:MDkAvG}) valid for a large number of domains.}
\label{fig:shSimplification}
\end{figure}

The effect of averaging and above discussed simplification [\eq{MDkAvG}] on $\MSq{\Dk{x}}$ is shown in the Fig.~\ref{fig:shSimplification}. This picture compares the square modulus of Fourier coefficients calculated for three different cases: the gray solid line depicts $\MSq{\Dk{x}}$ with  $\M{\Dk{x}}$  calculated from \eq{MDk}, red dashed line shows value of $\MSqAver{\Dk{x}}$ where $\MSq{\Dk{x}}$ is calculated as for grey line and then averaged over 300 randomly generated samples, finally the dotted blue line shows  $\MSqAver{\Dk{x}}$ calculated according to simplified \eq{MDkAvG}. One can see that results obtained with the simplified equation are consisted with numerically averaged SH intensity calculated using the full model for both very small (a) ($\sigma$=0.05 $\mu$m) and large (b) ($\sigma$=10 $\mu$m) dispersion. All curves were calculated assuming  number of domains to be  N=500 and average domains size $\lAver$=1 $\mu$m.

\section{Second Harmonic Generation}
We consider the situation where the fundamental beam with the wavelength $\lamFH$ is incident onto the sample along the $y$ direction (Fig. \ref{fig:Diagram}). Two photons of the fundamental harmonic $2\kFH$, generate non-collinear SH with wave vector $\kSH$ emitted at the angle $\eaSH$ such that the corresponding phase mismatch equals to $\DkSHvec=\kSHvec-2\kFHvec$ (triangle OAB in Fig.~\ref{fig:Diagram}). According to the formula \eq{ESF_2} (with $E_a=E_b=\EFH$) the amplitude of the generated SH electric field $\ESH$ is given as
\begin{equation}\label{eq:ESH} 
  \ESH(\DkSHvec{}) \propto i \frac{d L_y\EFH^2}{L_x}  \sinc\left(\frac{\DkSH_y L_y}{2}\right)\,\M{\DkSH_x}.
\end{equation}
The averaged SH intensity $\aver{\ISH}\propto\aver{|\ESH|^2}$ is:
\begin{equation}\label{eq:ISH}
  \aver{\ISH} \propto \frac{1}{L_x^2}\sinc^2\left(\frac{\DkSH_y L_y}{2}\right)\,\MSqAver{\DkSH_x}.
\end{equation}
\begin{figure}[th]
\centerline{\includegraphics[width=8.5cm]{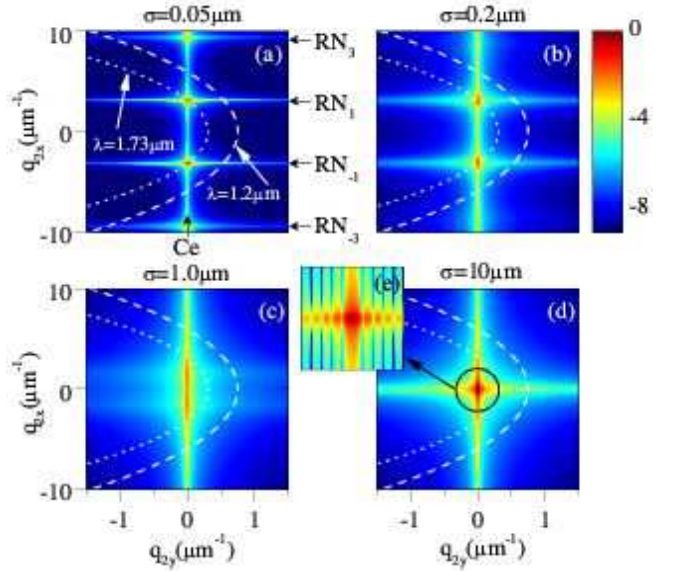}}
\caption{(color online) Illustrating the effect of randomness on the strength of SHG in nonlinear photonic crystal with random perturbation of the periodic  structure. Shown is the  map of the SH intensity (normalized to $\IZero$, logarithmic scale) in a Fourier space as a function of the phase mismatch $\DkSH{}_x$ and $\DkSH{}_y$. In all graphs the average domain size is $\lAver$=1 $\mu$m while the dispersion of the domain size varies from $\sigma$=0.05 $\mu$ m to $\sigma$=10 $\mu$m. The dashed and dotted  white curves represent all possible directions of the wave vectors of the second harmonics for  the fundamental wavelengths of  $\lambda$=1.2 $\mu$m and $\lambda$=1.73 $\mu$m, respectively. The inset depicts details of the map  in the vicinity of ($\DkSH_x=0, \DkSH_y=0$). Periodic oscillations of the intensity reflect the Makers fringes of the  forward emission.}
\label{fig:SHFourierSpace}
\end{figure}
\begin{figure}[th]
\centerline{\includegraphics[width=8.5cm]{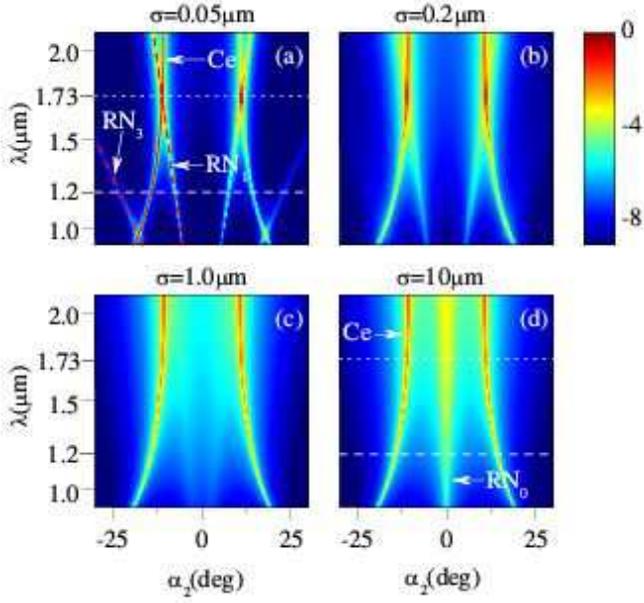}}
\caption{(color online) SH intensity (normalized to $\IZero$) as a function of emission angle $\eaSH$ and wavelength of the fundamental beam $\lambda$ for various degree of disorder: (a) $\sigma$=0.05 $\mu$m, (b) $\sigma$=0.2 $\mu$m, (c) $\sigma$=1 $\mu$m and (d) $\sigma$=10 $\mu$m.}
\label{fig:SHRealSpace}
\end{figure}
This formula relates the stochastic properties of the second order nonlinearity distribution to the spatial average intensity distribution  of the SH. The maximum SH intensity $\ISH$ is generated in case of perfect periodic structure and when SHG is transversely phase matched with first RLV (ie. $\DkSH_x=2\pi/\LAM$) and longitudinally phase matched (ie. $\DkSH_y=0$) for which intensity $\aver{\ISH}$ is:
\begin{equation}\label{eq:IZero}
  \aver{\ISH(\Dk{}_x=2\pi/\LAM,\Dk{}_y=0)}\equiv \IZero.
\end{equation}

The most universal way to analyze SH intensity is to use a two-dimensional map plotting the \eq{ISH} as a function of arbitrary phase mismatch components $\DkSH_x$ and $\DkSH_y$. This is demonstrated in Fig.~\rpict{SHFourierSpace} where we show SH intensity (normalized to $\IZero$) calculated for a sample with nominal domain width $\lAver$=1 $\mu$m and four different values of dispersion $\sigma$: plots (a-d) correspond to the increasing degree of the randomness ranging from  $\sigma$=0.05 $\mu$m (almost ideal periodic structure), $\sigma$=0.5 $\mu$m, $\sigma$=1 $\mu$m and $\sigma$=10 $\mu$m (strongly disordered structure). In case of periodic structure [Fig.~\rpict{SHFourierSpace}~(a)] there are five characteristic lines/traces. These  are: i) a vertical line at $\DkSH_y=0$ which  represents  the \v{C}erenkov SH radiation
\cite{Sheng:oe:2010,Tien:apl:1970,Sheng:11:Cerenkov} (marked as Ce), ii) two symmetrically positioned horizontal lines at $\DkSH_x$=$\pm$3.1 $\mu$m$^{-1}$ corresponding to first order Raman-Nath resonance~\cite{Saltiel:OL:09:848,Sheng:11:RN} (marked as RN$_{\pm1}$), iii) two symmetrically positioned horizontal lines at $\DkSH_x$=$\pm$9.4 $\mu$m$^{-1}$ corresponding to the third order Raman-Nath resonances (marked as RN$_{\pm3}$). When the dispersion of the domain size increases [(Fig.~\rpict{SHFourierSpace}~(b)] the  Raman-Nath emission lines shift towards the center and, at the same time, become broader and weaker until they disappear for strongly disordered structure (Fig.~\rpict{SHFourierSpace}~(c-d)). On the other hand the \v{C}erenkov SH signal can either increase or decrease depending on the value of $\DkSH_x$.

The dashed and dotted white curves in Fig.~\rpict{SHFourierSpace} represent the position of the phase mismatch vector $\DkSHvec$ calculated for two arbitrary chosen wavelengths of the fundamental wave: $\lambda$=1.2 $\mu$m (dashed line), and $\lambda$=1.73 $\mu$m (dotted line) using reported refractive index of LiNbO$_3$ crystal~\cite{Edwards:oqe:1984}. If we fix fundamental beam wavelength (say, $\lambda$=1.2 $\mu$m) it is clear that for almost periodic domain distribution the SH is efficiently generated only in those $\DkSHvec$ points where the white curve intersects  the  lines  corresponding to  large  value of the Fourier spectrum  of the nonlinearity modulation. Those characteristic points result in a wavelength dependent spatial emission pattern of SH.

Fig.~\ref{fig:SHRealSpace} shows the SH intensity $\aver{\ISH}$ (normalized to $\IZero$) as a function of fundamental wavelength and SH emission angle $\eaSH$. For a periodic structure [Fig.~\ref{fig:SHRealSpace}(a)] $\aver{\ISH}$ exhibits clear peaks where the emission angles $\eaSH$ are defined by partial or full phase matching conditions. For \v{C}erenkov SH emission, where $\DkSH_y=0$, the emission angle can be calculated as:
\begin{equation}\label{eq:SHCe}
  \cos(\eaSH^{Ce}) = \frac{n_1}{n_2},
\end{equation}
and for $m$-th order Raman-Nath where $\DkSH_x = m2\pi/\LAM$, emission angle is:
\begin{equation}\label{eq:SHRN}
  \sin(\eaSH^{RN_m}) = m\frac{\lamFH}{2\LAM n_2}.
\end{equation}
Those wavelength dependent emission angles are plotted in the Fig.\ref{fig:SHRealSpace}~(a) as a red solid line (marked as Ce) for $\eaSH^{Ce}$ and as red dashed lines (marked as RN$_1$, RN$_3$) for $\eaSH^{RN_m}$. As it has been mentioned earlier, it is clearly seen that the emitted Raman-Nath SH exhibits enhanced angular spreading for strong domain dispersion while \v{C}erenkov emissions is less sensitive to disorder.
\begin{figure}[tbh]
\centerline{\includegraphics[width=8.5cm]{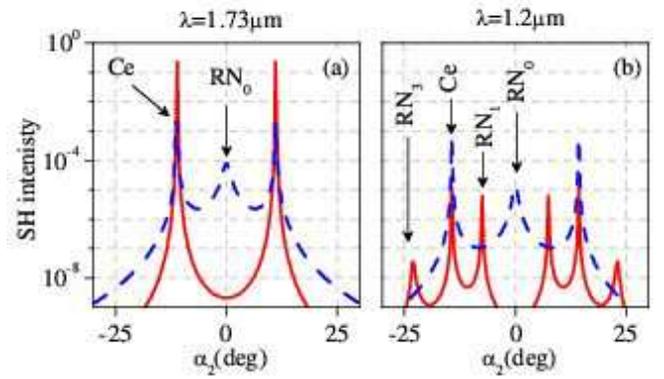}}
\caption{(color online) SH intensity (normalized to $\IZero$) as a function of the emission angle $\eaSH$ for $\sigma$=0.05 $\mu$m (solid red lines) and $\sigma$=10 $\mu$m (dashed blue lines) for two wavelengths: (a) $\lambda$=1.73 $\mu$m (cross section along dotted line in Figs.~\ref{fig:SHRealSpace}(a,d)) and (b) $\lambda$=1.2 $\mu$m (cross-section along dashed line in Figs.~\ref{fig:SHRealSpace}(a,d)).}
\label{fig:sHProfilesVsAlpha}
\end{figure}

The horizontal dashed and dotted lines in Fig.~\ref{fig:SHRealSpace} point to two arbitrary chosen wavelengths of the fundamental beam $\lambda$=1.2 $\mu$m (white dashed line) and $\lambda$=1.73 $\mu$m (white dotted line). The corresponding angular SH intensity profiles are depicted in Fig.\ref{fig:sHProfilesVsAlpha}, where red solid lines presents profile for  $\sigma$=0.05$\mu$m and blue dashed line for $\sigma$=10 $\mu$m. The strong modification of the emission pattern with high degree of randomness is evident. However it affects SH in different ways depending on the wavelength and the type of interaction. For $\lambda$=1.73 $\mu$m [Fig.\ref{fig:sHProfilesVsAlpha}(a)] we can see that increasing of $\sigma$ suppresses intensity of the \v{C}erenkov (Ce) signal. On the other hand for $\lambda$=1.2 $\mu$m [Fig.\ref{fig:sHProfilesVsAlpha}(b)] the \v{C}erenkov emission significantly increases while first and third order Raman-Nath peaks (RN$_1$ and RN$_3$) practically vanish. For both wavelengths strong dispersion produces zero order Raman-Nath (RN$_0$) emission.

\begin{figure}[tbh]
\centerline{\includegraphics[width=8.5cm]{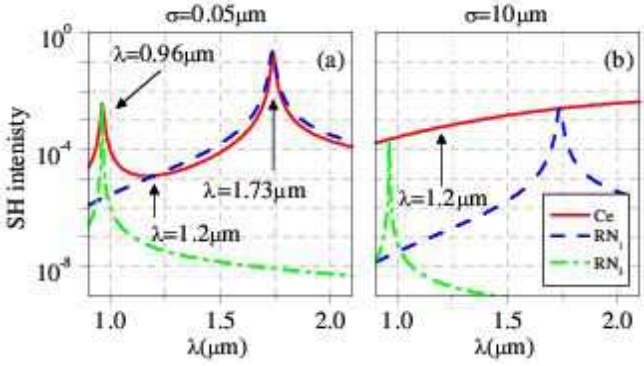}}
\caption{(color online) SH intensity (normalized to $\IZero$) of the \v{C}erenkov and Raman-Nath second harmonic signals as a function of fundamental wavelength for different dispersions of the domain size: (a) $\sigma$=0.05 $\mu$m and (b) $\sigma$=10 $\mu$m. Red solid line (Ce) represents \v{C}erenkov intensity, blue dashed line (RN$_1$) is the first order Raman-Nath and the green dashed-dotted line (RN$_3$) represents the third order Raman-Nath.}
\label{fig:sHProfilesVsLam}
\end{figure}
Fig.~\ref{fig:sHProfilesVsLam} shows \v{C}erenkov (red solid lines), Raman-Nath first (blue dashed lines) and third order (green dashed-dotted lines) SH as a function of wavelength for two different dispersions $\sigma$=0.05 $\mu$m [Fig~\ref{fig:sHProfilesVsLam}(a)] and $\sigma$=10 $\mu$m [Fig.~\ref{fig:sHProfilesVsLam}(b)].
One can notice that \v{C}erenkov radiation (Ce) is a subject to intensity modulation. The most efficient SH emission is expected where the Raman-Nath and \v{C}erenkov signals overlap (as it happens for $\lambda$=1.73 $\mu$m and $\lambda$=0.96 $\mu$m) what constitutes  the regime of the so called nonlinear Bragg diffraction~\cite{Berger:PRL:98}, i.e. SH emission under simultaneous fulfillment of the transverse and longitudinal phase matching. For strong disorder the peak intensity deteriorates due to decrease in RN intensity. However, at some wavelengths (as for $\lambda$=1.2 $\mu$m) the \v{C}erenkov  intensity increases even by 10$^2$ times. This strong dependence of the \v{C}erenkov signal on the wavelength is a result of the interference of many SH components emitted by each ferroelectric domain~\cite{Ren:arXiv:2010}. When the initially periodic domain pattern becomes disordered the interference effect gets weaker and the total \v{C}erenkov  signal is determined by the sum of the intensities of contributing waves. In effect, the \v{C}erenkov  signal which was strong for ceratin fundamental wavelength (due to a constructive interference) is weakened while the one  which was initially weak (due to destructive interference) increases.
\begin{figure}[htb]
\centerline{\includegraphics[height=4cm]{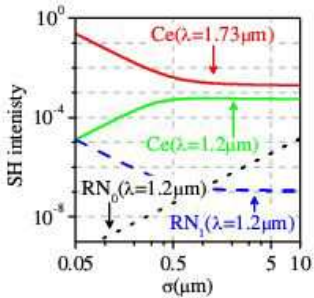}}
\caption{(color online) SH intensity (normalized to $\IZero$) of the \v{C}erenkov (Ce), zero (RN$_0$) and first (RN$_1$) order Raman-Nath signals as a function of the domain dispersion $\sigma$, for  $\lambda$=1.2 $\mu$m and $\lambda$=1.73 $\mu$m (as depicted in brackets).}
\label{fig:sHProfilesVsSig}
\end{figure}

This is even clearer in the Fig.~\ref{fig:sHProfilesVsSig} where we plot the \v{C}erenkov, zero and first order Raman-Nath intensities as a function of domain dispersion $\sigma$. We can see that with growing disorder Raman-Nath RN$_1$($\lambda$=1.2 $\mu$m) and \v{C}erenkov Ce ($\lambda$=1.73 $\mu$m) signals decrease and tends to a constant value as the constructive interference effects associated with periodicity of the domain distribution are  no longer relevant for large $\sigma$. On the other hand signal Ce($\lambda$=1.2 $\mu$m) increases due to weakening of destructive interference effect.
Zero order Raman-Nath RN$_0$ ($\lambda$=1.2 $\mu$m), initially not present for perfect structure ie. $\sigma\rightarrow0$, continuously increases with $\sigma$ as the high degree of disorder provides small but nonzero Fourier coefficients for $\DkSH_x$=0 which, in case of SH, contribute towards the forward emission.

\section{Third Harmonic Generation}
It has been demonstrated recently that nonlinear photonic crystals with random domain (or quadratic nonlinearity)
 distribution enable  third harmonic generation via cascading of the frequency doubling and sum frequency mixing~\cite{Molina:afm:2008,Wang:jpb:2010,Wang:oe:09,Sheng:apb:11}. In particular, it is shown that the cascading process can be completed via either \v{C}erenkov or Raman-Nath type phase matching \cite{Sheng:11:RN,Sheng:apl:11,Ayoub:apl:11}.

 In this section we provide the first analytical study of such process by extending the approach discussed in the previous two sections. We will consider the same setting of   1D  sample of the random nonlinear photonic crystal Fig.1(a) being illuminated by the  fundamental beam. The nonlinear interaction leads to simultaneous emission of the second and third harmonics. To calculate the amplitude of the third harmonic we use \eq{ESF_2} where $E_a=\ESH$ is taken from \eq{ESH} and $E_b = \EFH$. In this case the amplitude of the electric field of the third harmonic $\ETH$ is
\begin{equation}\label{eq:ETH}
 \begin{split}
   \ETH(\DkSHvec,\DkTHvec) & \propto - \frac{d^2 L_y^2 E_1^3}{L_x^2} \sinc\left(\frac{\DkSH_y L_y}{2}\right) \sinc\left(\frac{\DkTH_y L_y}{2}\right)\\
   & \times  M(\DkSH_x)\,M(\DkTH_x),
 \end{split}
\end{equation}
where $\DkSHvec$ and $\DkTHvec=\kTHvec - \kSHvec - \kFHvec$ represent the phase mismatch of the SHG and the sum frequency mixing, respectively. The phase matching diagram of the TH generation via cascading process is illustrated in Fig.~\rpict{Diagram}. This graph represents relevant processes in this multi-wave interaction. First a photon of SH is generated in the frequency doubling process with the phase mismatch wave vector of $\DkSHvec$ and emitted at the angle $\eaSH$ as depicted by the triangle OAB. Then this photon together with photon of the fundamental wave 
forms a photon
with the triple frequency of the fundamental  with the  phase mismatch vector $\DkTHvec$. The TH wave  is emitted in the direction determined by the angle $\eaTH$ (angle between $\vec{OF}$ and $\vec{OA}$). In order to reveal key features of this cascaded process we will first calculate the average intensity of the TH wave $\aver{\ITH}\propto\aver{|\ETH|^2}$ generated in such a single process:
\begin{equation}
 \begin{split}
  \aver{\ITH} &\propto \frac{1}{L_x^4} \sinc^2\left(\frac{\DkSH_y L_y}{2}\right) \sinc^2\left(\frac{\DkTH_y L_y}{2}\right) \\
    & \times \aver{\MSq{\DkSH_x}\MSq{\DkTH_x}},
 \end{split}
\end{equation}
where:
\begin{equation}
 \begin{split}
  &\aver{\MSq{\DkSH_x}\MSq{\DkTH_x}} = \frac{16}{\DkSH_x^2\DkTH_x^2} \\
    & \times\left(N^2\f(\DkSH_x)\f(\DkTH_x) + NO_1 + O_2\right),
  \end{split}
\end{equation}
where $O_1,\ O_2$ are function of phase mismatch $\DkSH_x,\DkTH_x$ and characteristic function $\ChDk{}$ and where $\M{\Dk{x}}$ was taken from \eq{MDkSimplified}. When $N\gg 1$ and $\left|\ChDk{}\right|<1$ terms containing $O_1$ and $O_2$ can be neglected and above simplifies to:
\begin{equation}\label{eq:ITH}
   \begin{split}
  \aver{\ITH} &\propto \left(\frac{4N}{L_x^2}\right)^2 \sinc^2\left(\frac{\DkSH_y L_y}{2}\right) \sinc^2\left(\frac{\DkTH_y L_y}{2}\right) \\
     & \times \frac{\f(\DkSH_x)\f(\DkTH_x)}{(\DkSH_x)^2(\DkTH_x)^2} = \aver{\ISF(\DkSHvec)}\aver{\ISF(\DkTHvec)}.
  \end{split}
\end{equation}
\begin{figure}[tbh]
\centerline{\includegraphics[width=8.5cm]{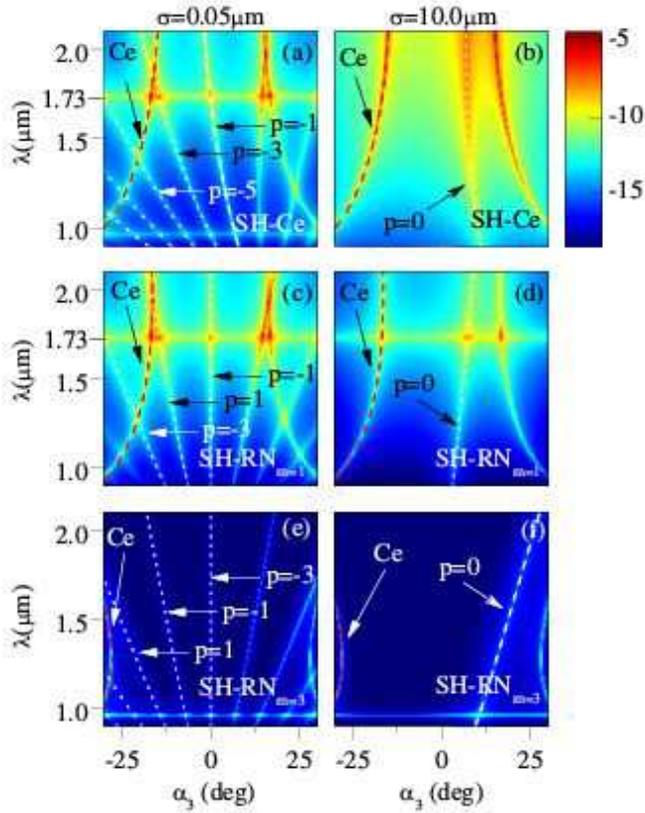}}
\caption{(color online) TH intensity calculated according to \eq{ITH} (normalized to $\IZero$) as a function of the wavelength of the fundamental beam and emission angle for very weak ($\sigma$=0.05 $\mu$m - left column) and strong ($\sigma$=10 $\mu$m - right column) domain dispersion. Pairs of graphs (a,b), (c-d) and (e-f) correspond to three distinct interaction mechanisms representing different type of the SH emission. They include the cases of second harmonic emitted as a \v{C}erenkov (a,b), as well as  first (c-d) and   third (e-f) order  Raman-Nath waves.}
\label{fig:THEmissionChanels}
\end{figure}
The above equation shows that the intensity of the third harmonic is a product of intensities resulting from  the constituent processes of SHG and the sum frequency mixing  utilizing the same Fourier space.

Since TH is generated in a cascaded process and its strength depends on the strength of the SH generated in a first step, it is natural to analyze how SH generated in different processes like \v{C}erenkov or Raman-Nath affects spatial TH emission. Fig.~\ref{fig:THEmissionChanels}(a) shows TH intensity $\aver{\ITH}$ calculated according to \eq{ITH} (normalized again to $\IZero$ given by \eq{IZero}) as a function of the fundamental wavelength and TH emission angle $\eaTH$ resulting only from SH generated in a \v{C}erenkov-like process, i.e. fixed values of SH phase mismatch vector to $\DkSH_y=0$ and $\DkSH_x=\DkSH\sin(\eaSH^{Ce})$. One can clearly observe a number of curves of enhanced TH intensity. Periodic structure exhibits strong resonances in Fourier space [see~Fig.\ref{fig:SHFourierSpace}(a)] and since both processes, SHG and following SFM utilize the same Fourier space it is expected to obtain high TH intensities for processes resulting from either \v{C}erenkov or Raman-Nath resonances~\cite{Sheng:11:RN}. The emission angle $\eaTH$ of TH generated in a double \v{C}erenkov process (ie when $\DkTH_y=0$) can be calculated as:
\begin{equation}\label{eq:THCeCe}
  \cos(\eaTH^{\text{Ce+Ce}}) = \frac{n_1}{n_3}.
\end{equation}
Angle $\eaTH^{\text{Ce+Ce}}$ as a function of wavelength $\lambda$ is plotted in Fig.~\ref{fig:THEmissionChanels}(a) as a dashed red line (marked as Ce) and it exactly overlaps with one of the TH signals. Similarly the remaining traces can be identified as interaction of Second Harmonic \v{C}erenkov  with following Sum Frequency Mixing of p-th order Raman-Nath (ie. $\DkTH_x=p2\pi/\LAM$) where the emission angle $\eaTH$ is:
\begin{equation}\label{eq:THCeRN}
  \sin(\eaTH^{\text{Ce+RN$_p$}}) = \frac{2 \LAM  \sqrt{n_2^2-n_1^2}+\lamFH p}{3 \LAM n_3}
\end{equation}
For the sake of clarity we plotted only few traces (white dashed lines): $p=-1$ , $p=-3$ and $p=-5$. The strong horizontal line at $\lambda$=1.73 $\mu$m and the weaker one at $\lambda$=0.96 $\mu$m represent  the enhanced SH generation at these wavelengths due to SH \v{C}erenkov resonance with RN$_1$ and RN$_3$ as shown in Fig.~\ref{fig:sHProfilesVsLam}.

When the domain structure becomes strongly disordered [Fig.\ref{fig:THEmissionChanels}(b)] all the previously observed Raman-Nath peaks disappear. However, as in the case of SH also now the \v{C}erenkov emission (red dashed line) remains relatively strong and additionally the zero order Raman-Nath $p=0$ is present (white dotted line).

Analogously Figs.~\ref{fig:THEmissionChanels}(c,d) show average intensity of the third harmonic $\aver{\ITH}$ but this time SH is generated as a first order Raman-Nath ($m=1$) and Figs.~\ref{fig:THEmissionChanels}(e-f) third order ($m=3$) Raman-Nath. The formulas for emission angles in those processes are:
\begin{equation}\label{eq:THRNCe}
  \cos(\eaTH^{\text{RN$_m$+Ce}}) = \frac{n_1 \pm n_2\sqrt{4 - m^2\frac{\lamFH^2}{\LAM^2n_2^2}}}{3n_3}
\end{equation}
for TH generated as a \v{C}erenkov (red dashed line) or
\begin{equation}\label{eq:THRNRN}
  \cos(\eaTH^{\text{RN$_m$+RN$_p$}}) = \frac{\lamFH (m + p)}{3\LAM n_3}
\end{equation}
when TH is generated as the $p$-th order Raman-Nath (indicated by white dotted line).

\begin{figure}[tbh]
\centerline{\includegraphics[width=8.5cm]{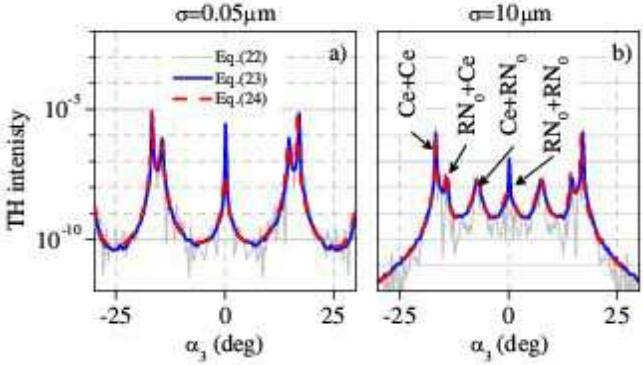}}
\caption{(color online) Comparison of TH generation via coherent process according to \eq{ITHCoher} (gray lines) and incoherent \eq{ITHIncoherent} (red lines) for a) small dispersion $\sigma$=0.05 $\mu $m and b) large dispersion $\sigma$=10 $\mu$m. The solid blue line represents coherent contributions to the overall THG  averaged over an ensemble of 120 randomly generated realizations of domain pattern [\eq{ITHCoherAver}]. Incident wavelength $\lambda$=1.73 $\mu$m. Intensity normalized to $\IZero$.}
\label{fig:THComparison}
\end{figure}
\begin{figure}[t]
\centerline{\includegraphics[width=8.5cm]{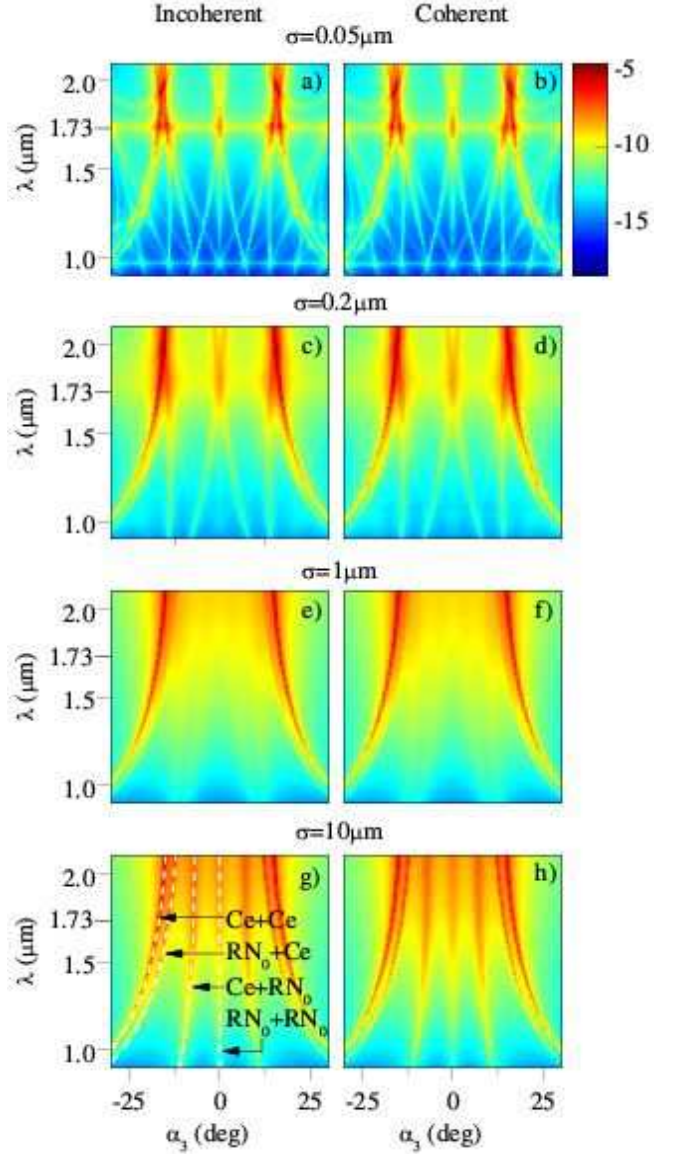}}
\caption{(color online) TH intensity (normalized to $\IZero$) as a function of the fundamental wavelength and emission angle in the regime of almost periodic ($\sigma$=0.05 $\mu$m) and disordered ($\sigma$=10 $\mu$m) nonlinear photonic crystal. Plots in the left column are obtained assuming incoherent contribution to the total TH intensity  according to \eq{ITHIncoherent}. Plots shown in  the right column are obtained by assuming coherent model  \eq{ITHCoherAver}, and by averaging the calculated TH intensity pattern  over 45 different realizations of the domain (nonlinearity) distribution.}
\label{fig:tHIncohTotal}
\end{figure}
In experimental reality there is no possibility to limit THG to only selected processes so, in general, the total intensity of the third harmonic $\ITH$ would be a sum of TH generated from all possible emitted second harmonics. As described previously TH can be generated in a process OBCF as schematically shown in Fig.\ref{fig:Diagram}. However, the TH with exactly the same wave vector $\vec{\kTH}$ can be also generated in an another process as, for example, the one indicated  by ODEF in the same figure. In order to calculate total TH emitted in a particular direction $\eaTH$ an integral over all possible SH emission angles $\eaSH$ has to be calculated $\ITH^{\text{coh}}\propto|\int\ETH\,d\eaSH|^2$ ie:
\begin{equation}\label{eq:ITHCoher}
 \begin{split}
  \ITH^{\text{coh}} &\propto \frac{1}{L_x^4} \Big|\int_{-\pi}^{\pi}\sinc\left(\frac{\DkSH_y L_y}{2}\right) \sinc\left(\frac{\DkTH_y L_y}{2}\right)\\
   & \times \M{\DkSHvec}\M{\DkTHvec}\, d\eaSH\Big|^2
  \end{split}
\end{equation}
and in case of presence of randomness its averaged version is given as
\begin{equation}\label{eq:ITHCoherAver}
 \begin{split}
  \aver{\ITH^{\text{coh}}} &\propto \frac{1}{L_x^4} \Big<\Big|\int_{-\pi}^{\pi}\sinc\left(\frac{\DkSH_y L_y}{2}\right) \sinc\left(\frac{\DkTH_y L_y}{2}\right)\\
   & \times \M{\DkSHvec}\M{\DkTHvec}\, d\eaSH\Big|^2\Big>,
 \end{split}
\end{equation}
where averaging again is performed over all realizations of domains widths. Both integrals \eq{ITHCoher} and \eq{ITHCoherAver} cannot be simplified and can be evaluated only numerically. However our previous experiments on third harmonic generation in nonlinear crystals with 2D fully random domain structure indicate that the emission process is in fact incoherent, i.e. the intensity of the TH is given as a superposition of intensities generated by all constituent second harmonic waves~\cite{Wang:jpb:2010,Wang:oe:09}. Therefore for large dispersion of domain sizes it is justified to calculate TH signal assuming incoherent emission $\aver{\ITH^{\text{inc}}} \propto \int\aver{\ITH}\, d\eaSH$ ie:
\begin{equation}\label{eq:ITHIncoherent}
 \aver{\ITH^{\text{inc}}} \propto \int_{-\pi}^{\pi}\aver{\ITH}\, d\eaSH = \int_{-\pi}^{\pi} \aver{\ISF(\DkSHvec)}\aver{\ISF(\DkTHvec)} \,d\eaTH
\end{equation}
where we use $\aver{\ITH}$ from \eq{ITH}.

In Fig.\ref{fig:THComparison} we show angular intensity profile of the third harmonic in case of two domain dispersions: $\sigma$=0.05 $\mu$m (a) and $\sigma$=10 $\mu$m (b) for fundamental wavelength $\lambda$=1.73 $\mu$m. The grey noisy line represents the TH intensity based on coherent superposition of contributing second harmonic waves, calculated for a single particular realization of domain distribution [\eq{ITHCoher}]. The solid blue line depicts coherently calculated intensity (same as grey line) but averaged over 120 realizations of randomly generated structures [\eq{ITHCoherAver}]. Finally, the dashed red line shows the intensity profile  obtained by assuming an incoherent interaction according to \eq{ITHIncoherent}. Both models clearly coincide confirming the incoherent character of third harmonic generation in that case. The labels next to the emission peaks in  Fig.\ref{fig:THComparison} identify processes responsible for particular emission angle. The labeling convention used here and in Fig.~\rpict{tHIncohTotal} is such that the first symbol refers to the second harmonic process while the second to the sum frequency mixing.

Fundamental wave $\lambda$=1.73 $\mu$m is special case when Raman-Nath peaks overlap with \v{C}erenkov. In order to ensure that the process is incoherent for broader spectrum we calculated the angular TH intensity distribution for wavelength range 0.9-2.1 $\mu$m. The results are shown in Fig.~\rpict{tHIncohTotal} which depicts the angular TH intensity distribution (normalized to $\IZero$) for few  values of the dispersions of the  domain width, $\sigma$=0.05 $\mu$m, $\sigma$=0.2 $\mu$m, $\sigma$=1 $\mu$m and   $\sigma$=10 $\mu$m. Maps in the left column are obtained  assuming incoherent contributions to the total intensity [\eq{ITHIncoherent}]. On the other hand,  maps in the right column represent coherent superposition of constituent harmonic waves (\eq{ITHCoherAver} averaged over 45 random samples). It is clear that both models lead to the same emission maps. While of course the actual intensity of the emitted TH signal in incoherent and coherent models will differ, in particular for small dispersion case, the angular dependence of the emitted TH  will be the same.
Again we clearly see that the out of many discrete TH emission channels clearly visible in ideal periodic structure only the strongest, namely those involving  \v{C}erenkov  and forward emission are preserved in highly disordered regime. The increase of the forward emission of TH is a direct consequence of the disorder-enhanced forward emission of the second harmonic as seen in  Fig.5(d) and Fig.6.

\section{Conclusions}
In conclusions, we investigated theoretically the second and third harmonic generation in 1D nonlinear photonic crystals with random distribution of nonlinearity.
  We derived analytical formulas describing emission properties of the second and third harmonics in the presence of domain disorder.  We showed that in the limit of large number of domains, the THG process is described by product of simple  expressions describing each of the constituent processes, i.e. SHG and sum frequency mixing.  We considered various processes involved in the frequency mixing and analyzed their properties as a function of wavelength and degree of disorder.  We show that  the randomness-induced incoherence in the wave interaction leads to deterioration and angular spreading of harmonic generated in the processes relying on  transverse phase matching such as Raman-Nath. On the other hand the \v{C}erenkov frequency generation is basically insensitive to the domain randomness.

\bigskip
This work was supported by the Australian Research Council.



\end{document}